\documentclass[10pt,conference]{IEEEtran} 
\usepackage{cite}
\usepackage{amsmath,amssymb,amsfonts}
\usepackage{algorithmic}
\usepackage{graphicx}
\usepackage{textcomp}
\usepackage{xcolor}
\usepackage{cite}
\def\BibTeX{{\rm B\kern-.05em{\sc i\kern-.025em b}\kern-.08em
		T\kern-.1667em\lower.7ex\hbox{E}\kern-.125emX}}
\usepackage{framed}
\usepackage{url}
\usepackage{pdfpages}
\usepackage{listings}
\lstnewenvironment{JAVA}{\lstset{language=Java}}{}
\lstnewenvironment{JAVAN}{\lstset{language=Java,numbers=left}}{}

\newenvironment{iquote}{\bgroup\itshape\begin{quote}}{\end{quote}\egroup}

\begin{document}
	
	\title{How do students test software units?}
	
	\author{
		\IEEEauthorblockN{Lex Bijlsma}
		\IEEEauthorblockA{\textit{Department of Computer Science} \\
			\textit{Open Universiteit}\\
			Heerlen, Nederland\\
			lex.bijlsma@ou.nl}
		\and
		\IEEEauthorblockN{Niels Doorn}
		\IEEEauthorblockA{
			\textit{Academy of ICT \& Creative Technologies}\\ \textit{NHL Stenden Hogeschool}\\
			Emmen, The Netherlands \\
			niels.doorn@nhlstenden.com}
		\and
		\IEEEauthorblockN{Harrie Passier}
		\IEEEauthorblockA{\textit{Department of Computer Science} \\
			\textit{Open Universiteit}\\
			Heerlen, The Netherlands\\
			harrie.passier@ou.nl}
		\and
		\IEEEauthorblockN{Harold Pootjes}
		\IEEEauthorblockA{\textit{Department of Computer Science} \\
			\textit{Open Universiteit}\\
			Heerlen, The Netherlands\\
			harold.pootjes@ou.nl}
		\and
		\IEEEauthorblockN{Sylvia Stuurman}
		\IEEEauthorblockA{\textit{Department of Computer Science} \\
			\textit{Open Universiteit}\\
			Heerlen, The Netherlands\\
			sylvia.stuurman@ou.nl}
	}
	
	\maketitle
	
	\begin{abstract}
		We gained insight into ideas and beliefs on testing of students who finished an introductory course on programming without any formal education on testing.  We asked students to fill in a small survey, to do four exercises and to fill in a second survey.		
		We interviewed eleven of these students in semi-structured interviews, to obtain more in-depth insight. 		
		The main outcome is that students do not test systematically, while most of them think they \emph{do} test systematically.
		One of the misconceptions we found is that most students can only think of test cases based on programming code. Even if no code was provided (black-box testing), students try to come up with code to base their test cases on. 	
	\end{abstract}
	
	
	\section{Introduction}\label{introduction}
	Professional software developers spend a considerable part of their time on testing.
	Agile methods in particular have increased the importance of testing throughout the development process.
	Yet, many recent graduates lack sufficient testing skills~\cite{Pham-2017}.
	For example, Edwards found that students detected only approximately 15\% of the bugs, and observed that students often only apply so-called  `happy path' testing~\cite{Edwards-2014}, implicitly assuming that the input is `ideal'.  
	
	In most university curricula, scant attention is paid to testing.
	For example, testing is fully integrated within the curriculum in only 2 out of 20 Dutch universities~\cite{Doorn-Thesis}.
	In some curricula, testing is not a topic at all.
	In most curricula, testing education is limited to an introduction of Java's popular test framework JUnit.
	How to compose test cases is often not given explicit attention. 

	There are strong indications that testing instruction influences the quality of students' programs positively.
	Some educators state that testing as activity improves software comprehension~\cite{Edwards-2004} and use that motivation to introduce testing and integrate testing in programming education. 
	In fact, knowledge about testing in itself tends to improve the quality of students' programs~\cite{Lemos-Impact}.
	This effect was even observed when test cases were provided by the teacher~\cite{Brito-Experience}.
	Moreover, testing is one of the core knowledge areas in both the \textsc{acm/ieee} curriculum guidelines for computer science and software engineering~\cite{ACM-2013,ACM-2014}.
	
	When we want to produce explicit, procedural guidance on how to create test cases and better instructional material, we should know how students view testing initially and which misconceptions need to be paid attention. 
	Our research question is, therefore,
	\begin{itemize}
		\item What ideas do students have about testing before they have had any relevant instruction?
	\end{itemize}
	
	We provide an overview of what is known with respect to our question in Section~\ref{sec:Related}.
	In Section~\ref{sec:Method} we explain how we approached our research.
	The results are described and analyzed in the sections~\ref{sec-pre} through \ref{subsec:analysis_of_interviews}.
	We conclude with answers to our question.
	
	\section{Related work}\label{sec:Related}
	As far as we know, only Edwards systematically examined how students test. 
	He used the tests that students sent in for an assignment in a course on data structures~\cite{Edwards-2014}. 
	Grading was based on branch coverage of the tests. 
	The mean coverage was 95.4\%. 
	The similarity between the tests was large: 90\% of the tests were the same. 
	To check which bugs the student's tests could detect, all tests of the students were combined into a single large test suite, along with the reference tests for the assignment. 
	This test suite was then run against all student programs. 
	The tests of the students only detected 13.6\% of the total number of bugs. 
	Almost all students only performed `happy path' testing, testing only the default scenario. 
	
	The fact that not only students, but also professionals tend to rely on `happy path testing' has been known for a long time. Leventhal~\cite{Leventhal-1994} found strong evidence of happy path testing (also called positive bias testing). The only `antidote' mentioned here  is to construct thorough and complete program specifications. 
	
	The tendency of students only to test the default scenario is in accordance with the finding that students have `alternative standards' for correctness~\cite{Kolikant-2005}. 
	Students soften the requirement that a function should show correct behavior for \emph{all} input, to the notion that the behavior should be correct for \emph{most} input, for input that seems `logical'. 
	
	Practitioners complain that many recently graduated students lack sufficient testing skills~\cite{Pham-2017}. 
	Practitioners see a skill gap between university graduates and industry expectations and think that graduates often do not seem to see the value of testing. 
	They observe that graduated students often follow a trial-and-error strategy: build something and `see if it works'.
	
	Explicitly teaching testing does have effect: students who have been educated in the subject produce better test cases~\cite{Gomez-Impact}.
	
	Testing, so it seems, requires explicit attention in the curriculum. 
	Without an explicit specification given by the teacher, students seem to assume a specification that only allows `ideal' input.
	
	\section{Method}\label{sec:Method}
	\subsection{Aim of the study}
	This is an in-depth explorative study on the perceptions of software testing by first-year students in computer science. 
	The emphasis is on qualitative data.
	The participants have programming knowledge on the level of an introductory course about programming, but have no prior formal education on the topic of software testing.
	We want to study their natural way of testing software units during programming.
	To do so, we used surveys, exercises and interviews.
	In this section, we describe our research setup, methods and the way we analyzed the results.
	
	\subsection{General approach}
		Thirty-one students were involved, all first-year computer science students at a university of applied science. 
	All students have basic knowledge of:
	\begin{itemize}
		\item \textsc{html}, imperative programming using \textsc{php} (period 1).
		\item Databases and \textsc{sql}, with some attention to exception handling (period 2).
		\item Introduction to \textsc{oo} Programming with Java using BlueJ (period 3).
	\end{itemize}
	In none of these courses testing was a topic. 
	
	We asked the students to fill in a survey. Then we asked them to do four exercises, and then we asked them to fill in a second survey.
	Together, this took them about 45 minutes.
	Finally, we interviewed a subset of the participants. 
	These interviews took about twenty minutes. 
	
	\subsubsection{Pre-exercises survey}
	The aim of the first survey is to test the ideas and beliefs the participants have on software testing without being exposed to the exercises.
	This survey contains one multiple-choice question about when during development tests can be best formulated, followed by an open question to motivate the given answer, and 
	six statements about testing with five-point Likert scale answer options, three of which were taken from Kolikant~\cite{Kolikant-2005}.
	
	\subsubsection{The exercises}
	The participants were presented with four exercises. 
	Three exercises had a functional description and implementation. One had a functional description only.
	For each exercise, the participant is asked:
	\begin{itemize}
		\item to give the test cases needed to decide whether the function is correct or not
		\item (for the first three) to determine the correctness of the provided implementation, and, in case of incorrectness, to provide a test case to prove this claim.
	\end{itemize}
	
	\subsubsection{Post-exercises survey}
	The second survey is held directly after the exercises. 
	The survey asks for:
	\begin{itemize}
		\item the perceived complexity of the exercises
		\item the process they followed to complete the exercises
		\item the supposed correctness of their answers
		\item the general ratio of time that they think should be spent on programming and testing in daily practice
		\item whether they took boundary values into account during the exercises.
	\end{itemize}
	All questions of this survey have five-point Likert scale answer options.
	
	\subsubsection{Interviews}
	To obtain more in-depth knowledge of the way the exercises have been done, we interviewed a subset of the participants using a semi-structured interview. 
	We could also verify the given answers to the exercises and the students' ideas about testing in more depth.
	All interviews were conducted by two interviewers, one as the chair and the other one making notes.
	All interviews were audio-recorded and written down verbatim.  
	
	\subsection{Ideas about testing}
	We aimed to study both the student's testing methods and the way they understand the concepts related to testing. 
	We also gained insights into the misconceptions of the participants.	
	We discerned the following ideas about testing. The instruments we used are in braces:
	\begin{itemize}
		\item During what programming phase (before, during or after programming) is testing relevant (pre-exercises survey)?
		\item Which stakeholder should conduct testing? (pre-exercises survey)
		\item In what depth and width should software be tested (pre- and post-exercises surveys, exercises and interviews)?
		\item What time ratio should be spent on testing? (post-exercises survey)	
		\item Completeness of testing, i.e. when does one have enough test cases (post-exercises surveys, exercises and interviews)?
		\item Does creating test cases help with understanding code (pre- and post-exercises surveys)?
		\item Does one use boundary values for test cases (pre- and post-exercises surveys)?
	\end{itemize}

	We also asked them to note the start and end time at each exercise, to see how much time they spent creating test cases.   
	
	\subsection{Analysis}
	The analysis was done by four researchers, all involved in software engineering education.   
	
	\subsubsection{Pre- and post-exercises surveys}
	Both surveys were subjected to quantitative analysis. 
	
	Answers to the open question from the pre-exercises survey were collected and analyzed quantitatively.
	We labeled the answers using characteristics that we found in the answers.
	Examples of these characteristics are:
	\begin{itemize}
		\item before, during or afterwards testing
		\item testing the whole by testing the individual components
		\item iterative approach: write some code, immediately write tests
		\item trial and error approach
		\item focus on code
		\item focus on functionality
		\item testing as a means to check whether the code is robust.
	\end{itemize}
	The analysis was performed by two researchers and reviewed by the other two researchers.
	
	\subsubsection{Exercises} 
	The answers to the exercises were analyzed separately on completeness of the test cases, test approach, mistakes, misconceptions, and time spent. 
	After that, the findings were aggregated using a classification, defined during a brown paper session. 
	The results of these findings and classification were discussed until consensus of all decisions was reached.
	
	\subsubsection{Interviews} 
	All transcripts and notes were read entirely by all researchers.
	We analyzed the interviews with respect to the completeness of the test cases and the approaches. 
	We used a classification, defined during a brown paper session, to aggregate the results and analyzed them quantitatively. 
	The results of these findings and classification were discussed until we obtained consensus of all decisions.
	
	\subsubsection{Meta-analysis}
	Finally, to determine the main findings, we performed a meta-analysis.
	We determined the most important classes during a brown paper session, and provided them with clear examples. These are the examples we show in this article.
	Again, the results of this analysis and classification were discussed until we obtained consensus of all decisions.
	
	\section{Results - Pre-exercises survey}\label{sec-pre}
	31 students filled in the survey.
	
	\subsection{When to test}\label{when-to-test}
	Question: \textit{The best time to construct test cases is (a) after, (b) before or (c) during programming? Motivate your answer. (Multiple answers are allowed.)} 
	
	Answers: 
	
	\begin{tabular}{|l|l|}
		\hline
		combinations & frequency ($N=31$)\\
		\hline
		a & 2 \\
		b & 2 \\
		c & 12 \\
		ab & 0 \\
		ac & 9 \\
		bc & 1 \\
		abc & 5\\
		\hline
	\end{tabular} \\
	
	If we just look at the number of times an alternative was mentioned, whether or not in combination with other alternatives, we get the following:\\
	
	\begin{tabular}{|l|l|}
		\hline
		alternative & frequency ($N=31$) \\
		\hline
		after & 16 \\
		before & 8 \\
		during & 27\\
		\hline
	\end{tabular} \\ 
	
	Conclusion: There is a preference for testing during programming. 	
	The low score for testing before programming is understandable if students base their test cases on code inspection (see later). 
	We also suspect, based on the motivations students added, that some students do not clearly distinguish between constructing test cases and running tests. 
	Also, the preference for testing during programming is possibly based on a confusion between testing and compiling, and between running and debugging (see later).
	
	\subsection{Claims about testing}\label{ssc:Claims}
	We presented the students with six statements about testing. The questions 4, 5 and 6 were taken from Kolikant~\cite{Kolikant-2005}. 
	The students could indicate the level of agreement on a five-point Likert scale (1: completely disagree, 5: completely agree). For these statements, we have $N=29$.
	
	\begin{enumerate}
	\item Absence of errors\\
	Claim: \textit{Testing can make it plausible that your program does not contain errors.}
	
	Answers: Average 3.41, standard deviation 1.02. 
	Many students place high trust in the power of testing.
	
	\item Who tests?\\
	Claim: \textit{It is best if end users perform the tests.}
	
	Answers: Average 3.66, standard deviation 1.20. 
	This statement is widely agreed with, which is unexpected in view of the preference they expressed to construct test cases during development, see~\ref{when-to-test}.
	
	\item Which test cases?\\
	Claim: \textit{The most important consideration when selecting test cases is to ensure that they are representative of the expected use of the program.}
	
	Answers: Average 3.59, standard deviation 1.05. 
	This shows the tendency to `happy path testing'.
	
	\item Confidence\\
	Claim: \textit{For a program I have written myself, I know it works well when I have run it several times and obtained correct output.}
	
	Answer: Average 2.66, standard deviation 0.98. 
	This is similar to claim A.1 from Kolikant's paper~\cite{Kolikant-2005}. 
	In that paper, 50\% of respondents agreed with the statement, both at high school and college level. 
	Our respondents seem to possess a somewhat more sophisticated attitude: only 24\% agreed. 
	
	\item Reasonable output\\
	Claim: \textit{In testing a program for a complicated calculation, I am satisfied when the output looks reasonable. It is not necessary to redo the calculation by hand.}
	
	Answer: Average 1.69, standard deviation 0.70. This is similar to claim A.2 from Kolikant~\cite{Kolikant-2005}. 
	33\% of the high school students and 69\% of the college students agreed in his study.
	Of our respondents, only one agreed, choosing answer 4. 
	
	\item No testing\\
	Claim: \textit{Sometimes I am sure that a program I have written is completely correct. In such a case, if the program compiles, it is not necessary to run or test the program.}
	
	Answer: Average 1.55, standard deviation 0.90. 
	This is similar to claim A.3 from Kolikant~\cite{Kolikant-2005}. 
	In the Kolikant study, 42\% of high school students and 31\% of college students agreed. 
	Of our respondents, only two (7\%) agreed. Both gave the answer 4.
	\end{enumerate}
	
	\section{Results - Exercises}\label{sec:analysis_of_exercises}
	Students were presented with four exercises. 
	In each exercise, they were asked to write test cases for the given function.
	Three of the exercises were both `black-box' and `white-box': both a functional specification and Java code were provided.
	One exercise was `black-box' with only a functional specification.
	
	For each of the white-box exercises, students were asked whether they considered the code to be correct.
	If not, they were asked to present a test case that would fail due to the incorrect code. 
	
	All exercises were single functions with input and output in the form of arrays of integers or a single integer.
	The exercises contained programming constructs and syntax that should be familiar to the students and were part of the previous Java courses they followed. 
	
	\subsection{The exercises}
	The exercises were as follows:
	
	\subsubsection{Exercise 1: The longest period of frost}
	This function determines the length of the longest period of frost from a series of temperatures. 
	The input is an array of integers representing the temperatures of a sequence of days. The output is an integer representing the length of the longest number of consecutive days the temperature was below zero.
	The body of the method is not correct: \texttt{currentPeriod} is initialized to -1, which should be 0.
	The code is as follows:
	
	\begin{lstlisting}[caption={Frost exercise (frost)}, captionpos=b]
	/**
	* Returns the longest uninterrupted period of temperatures below 0
	*/
	public int longestBelowZero(int[] temperatures) {
		int longestPeriod = 0;
		int currentPeriod = -1;
		for (int i=0; i < temperatures.length; i++) {
			if (temperatures[i] < 0) {
				currentPeriod++;
			} else {
				if (currentPeriod >= longestPeriod) {
					longestPeriod = currentPeriod;
				}
				currentPeriod = 0;
			}
		}
		return longestPeriod;
	}
	\end{lstlisting}
	
	\subsubsection{Exercise 2: The lowest index of the lowest value}
	The input for this function is an array of integers. The function should determine the lowest index of the lowest number in the array. 
	The provided code is incorrect: the index in the for loop that iterates over the values is initialized to \texttt{1}, which should be \texttt{0}. 
	The code is as follows:
	
	\begin{lstlisting}[caption={Lowest index of the Lowest value exercise (min-min)},captionpos=b]
	/**
	* Returns the lowest index of the lowest value
	*/
	public int findTheLowestIndexOfTheLowestValue(int[] numbers) {
		int index = 1;
		for (int i = 1; i < numbers.length; i++) {
			if (numbers[i] < numbers[index]) {
				index = i;
			}
		}
		return index;
	}
	\end{lstlisting}
	
	\subsubsection{Exercise 3: Changing coins}
	The input of this function is an integer representing an amount of money in cents. The function returns the smallest sequence of coins from the euro coin series that can be used to represent that amount of money. 
	The code is correct.
	
	\begin{lstlisting}[caption={Exchange exercise (coins)},captionpos=b]
	/**
	* Returns the smallest sequence of coins  
	*    to represent the input argument.
	* Possible coins: 
	* 1, 2, 5, 10, 20 and 50 cent and
	* 1 and 2 euro (100 and 200 cent)
	*/
	public ArrayList<Integer> exchange(int amount) {
		ArrayList<Integer> result = new ArrayList<>();
		int[] coins = {200, 100, 50, 20, 10, 5, 2, 1};
		for (int coin : coins) {
			for (int i=0; i < amount / coin; i++) {
				result.add(coin);
			}
			amount = amount - (amount / coin) * coin;
		}
		return result;
	}
	\end{lstlisting}
	
	\subsubsection{Exercise 4: Palindrome} 
	The input of this function is a string. It returns true when the input is a palindrome, and false when it is not. 
	This is the black-box exercise, without implementation. Only the signature and the description are given.
	
	\begin{lstlisting}[caption={Palindrome checker exercise (palindrome)},captionpos=b]
	/**
	* Input: A string
	* Output: true if the string is a palindrome
	*        otherwise false
	*/
	public boolean isPalindrome(String word) {
		// no body is provided
	}
	\end{lstlisting}
	
	These four exercises have an algorithmic nature. 
	Based on the description and the code, students should be able to understand the algorithm and come up with test cases.
	The exercises differ in the concepts that are used. 
	The frost exercise contains a for-loop that iterates over the input array. 
	It also has branching, a conditional statement (\texttt{if else}), with another conditional statement in the \texttt{else} branch. 
	The second exercise uses the same array with two indexes in the conditional statement. 
	This can be easily overlooked. 
	The coin exchange exercise uses a nested loop construct. 
	This is often considered to be a complex concept for novice programmers~\cite{Cetin2015,Ginat2004}.
	There is also a mathematical statement with a subtraction, a multiplication and a division. 
	The palindrome exercise handles string input and uses a boolean return value. 
	
	\subsection{Observations}
	Analyzing the students' answers, we divided our observations into four main categories: 
	\begin{enumerate}
		\item test approaches
		\item completeness of the test cases
		\item misconceptions
		\item programming knowledge
	\end{enumerate}
	For each category, we give some examples. 
	The exercise is mentioned between brackets, i.e. frost, coins, min-min, or palindrome.
	
	\subsubsection{Test approaches}\label{subsec:TestApproachesApplied}
	
	\paragraph{Happy path testing}
	The test approach that most students applied is happy path testing. 
	We determined 33 test sets consisting of happy path test cases only: frost 10, coins 3, min-min: 6, and palindrome: 14. 
	An example from the frost exercise:
	
	\begin{iquote}`One test case with at least one temperature below zero.'\end{iquote}
	
	\noindent
	Another example from the min-min exercise:
	\begin{iquote}`[0,1,2,0,2,-1]'\end{iquote}	
	\paragraph{Structural approaches}
	We determined nine test cases that could be interpreted as boundary testing.
	
	One student differentiated on the frost exercise between a test case with one temperature and a test case with several temperatures:  
	\begin{iquote}`One test with a known outcome. Then, one array with one item. And one array with temperatures below zero only.'\end{iquote}
	\noindent
	Another student defined an empty array and a number of arrays with length greater than zero for the frost exercise:
	\begin{iquote}`[], [-1,-1,-1,4,3,-2,-2], [-1,0,1,2,3,], [-1,-1,2,2].'\end{iquote}
	\noindent
	One student applied a more or less systematic test approach. 
	The student described four test cases (frost):
	\begin{iquote}`An array without temperatures below zero,
		an array with one period of frost,
		an array with multiple periods of frost,
		and an array with a period of frost at the beginning.'\end{iquote}
	\noindent
	The last test case shows the bug in the code. Nevertheless, the students did not define test cases with array's of length zero and one.
	
	\paragraph{Test cases based on code inspection — bug find}
	In four situations a student only wrote one test case based on the bug  in the code. One example (frost):
	\begin{iquote}`[-1,-1,0,1]'\end{iquote}
	\noindent The function's output is 1, while the longest period of frost is 2, due to the wrong initialization of \texttt{currentPeriod}. 

	\noindent
	Another example (min-min):
	\begin{iquote}`[\dots] after that I would use a test case where the lowest value is on the first index, this will probably fail because of the for-loop which starts with \texttt{i=1}. Personally, I would probably never test this because I would have noticed this while programming.'\end{iquote}
	
	\noindent
	This test case demonstrates the bug. The last part of the quote underlines the approach of code inspection as an alternative for testing. 
	
	One student mentioned that the body of the change function (coins) is incorrect, but was not able to define a test case showing the bug.
	
	\paragraph{Miscellaneous}
	One student gave an answer we do not understand (frost); it might be an approach to debugging instead of testing:

	\begin{iquote}`You have to see the array to figure out if it is correct.'\end{iquote}
	
	\noindent One student was unable to provide a concrete test case. 
	This student gave the following description (frost): 
	
	\begin{iquote}`An input value of which you know the output value'\end{iquote}
	
	\noindent
	which is basically a very high-level description of testing in general.
	
	\subsubsection{Completeness of test cases}
	A complete test set should discern several aspects, for instance, structural as well as domain-specific, or specification as well as implementation based test cases. 
	Implementation based test cases are only possible, of course, if an implementation is present~\cite{bijlsma2018}.
	
	In case of the frost exercise, examples of structural aspects are an empty array, an array with one element and an array with several elements.
	Examples of domain-specific aspects are no frost periods at all and frost periods of different lengths spread out over the array in several ways. 
	An example of an implementation aspect is what to do in case of an anomaly. 
	If the implementation is present, one can think of applying various coverage criteria as well as testing overflow situations in cases specific types are used for variables. 
	
	We observed that almost all the test sets defined by the students are far from complete.
	For example, for the frost exercise, a minimum of three test cases is needed to have path coverage.
	Only one student provided enough test cases to reach path coverage, as follows:
	\begin{iquote}`[0,-1,1,-1,0,-1], [-1,0,-1,-1], [1,1,0,1]'\end{iquote}
	Most students defined either one or two test cases, or they provided  test cases that could not test the given functions sufficiently.
	For example, for the first exercise, we found only one test case eleven times.

	\begin{iquote}`One test case with at least one temperature below zero.'\end{iquote}
	
	\noindent We found only 2 test cases 4 times
	\begin{iquote}`One test case with a negative number and one test case with a positive number.'\end{iquote}
	 
	We found three test cases 4 times  and four test cases 2 times. 
	As mentioned before, most of these test cases test the happy path scenario only.
	
	\noindent An example of an incomplete test case for the coin exercise:
	\begin{iquote}`Test cases with multiples of coins.'\end{iquote}
	
	The students who applied a more systematic testing approach had a slightly more complete test set. 
	
	\subsubsection{Misconceptions}
	
	\paragraph{Exhaustive testing}
	One student presented one test case and then proposed an exhaustive testing scenario (frost):
	\begin{iquote}`[-1,-1,0,0,1,1] and I shall look to all possible inputs and see whether the program reacts as is expected with several days of frost.'\end{iquote}
	
	\paragraph{Test cases without expected result} 
	Some students specified an array with random numbers as a test case, for example
	\begin{iquote}`Random numbers in an array.'\end{iquote}
	\noindent
	The problem with this approach is, of course,  that the result of such a test case is unknown beforehand and therefore it is impossible to determine the correctness of the function. 
	
	\noindent
	Another example (min-min) also defining a random array as test case is: 
	\begin{iquote}`1.-) One array with equal numbers, 2.-) one array starting with the lowest number, 3.-) one array with all random numbers, and 4.-) one array with numbers you know the result of.'\end{iquote}
	
	\paragraph{Type testing}
	One student defined a test case with an array containing a character, where an array with integers is expected (min-min):
	
	\begin{iquote}`One array with two numbers, one array with a lowest number, and an array with a character.'\end{iquote}
	\noindent
	The language used is Java, a strongly typed language.
	
	\paragraph{Dividing by zero}
	Two students remarked that dividing by zero is forbidden and thought that, as a consequence, dividing zero by something else is forbidden as well.
	
	\paragraph{Implementation is required}
	As part of the palindrome exercise, one student wrote one test case (`lol', which is a palindrome), but mentioned that it is impossible to check the case because the implementation is missing.
	
	\subsubsection{Programming knowledge}
	Although these students should have the required knowledge about Java, it seems that some students struggle with the given code. 
	For example, one student wrote as a test case (coin):
	\begin{iquote}`49,7,9,127,61 I think that something is missing with `int coin', because an int can not be an array.'\end{iquote}
	Here, this student, probably, has not enough knowledge of Java types. 
	
	Some students are not focused on input-output testing, but on print-based testing to check whether certain statements are successfully executed and in what order.  
	For example, one student wrote as a test case (min-min):
	\begin{iquote}`I define a method that prints the array to see the array is successfully created.'\end{iquote}
	
	One student had no idea how to solve this exercise (frost) and stated:
	\begin{iquote}`I have no idea.'\end{iquote}
	
	One student did not understand the palindrome exercise, judging by the answer:
	\begin{iquote}`droom, paling, moordnilap, true, false, 12345, palindr00m'\end{iquote}
	
	\subsubsection{Time spent on the exercises}
	The students were asked to note the start and end time for each exercise. We got the following averages per exercise:\\
	
	\begin{tabular}{|l|l|l|}
		\hline
		Exercise & Average time spent & standard deviation\\
		\hline
		1 ($N=31$) & 6:42 & 3:07 \\
		2 ($N=30$) & 4:30 & 1:46 \\
		3 ($N=31$) & 4:54 & 1:59 \\
		4 ($N=31$) & 2:17 & 1:16 \\
		\hline
	\end{tabular}\\
	
	The first exercise took the students the longest.
	This could be caused by the time needed to understand how the exercises worked.
	During the interviews, students did consider the third exercise to be the most difficult. 
	The last exercise, the black-box exercise, took considerably less time than the exercises with the code provided.
	This supports our findings that students mainly use the code to think of test cases.
	
	Overall, the short time spent by students to solve these exercises strikes us. 
	This finding matches with the findings of happy path testing, and test cases based on code inspection.
	
	\subsubsection{Correct or incorrect}
	For each white-box exercise, students were asked if the code was correct and if not, to come up with a test case to support their claim. 
	Exercise one and two both contained one logical error and exercise three was correct. None of the exercises contained syntax errors. 
	The following table shows the results:\\
	
	\begin{tabular}{|l|l|l|l|}
		\hline
		Exercise & Correct & Incorrect & Valid test case \\
		\hline
		1 ($N=22$) & 11 & 11 & 7 \\
		2 ($N=25$) & 4 & 21 & 19 \\
		3 ($N=23$) & 10 & 13 & n/a \\
		\hline
	\end{tabular}\\
	
	With respect to the first exercise, an equal number of students thought that the code is correct and incorrect. 
	Seven students were able to provide a valid test case to support their claim. 
	On the second exercise, students scored a lot better. Most students noticed the bug and were able to provide a test case to support their claim. 
	With the third exercise, most students wrongly think the code is incorrect.
	This supports the indication that most students found this exercise the hardest.
	
	\section{Results - Post-exercise survey}\label{sec:survey}
	The survey has been filled in by 31 students. The following statements were submitted to the students after they had performed the exercises.
	
	\subsection{Understanding}
	Claim: \textit{Having to think of test cases has increased my understanding of the program code.}
	
	Answer: Average 3.42, standard deviation 1.13. We did not verify whether any deeper understanding was actually reached, but students feel they did reach an increased understanding.
	
	\subsection{Test coverage}
	Claim: \textit{My test cases were sufficient to test the program.}
	
	Answer: Average 3.10, standard deviation 0.83. In fact the test cases were clearly insufficient, which the students only realized when discussing them during the interview phase. Apparently, many students interpreted the exercise as `find the coding error in this program' and stopped when they had found one.
	
	This attitude could have been stimulated by the question to consider the correctness of the code, and if not correct, to present a test case that would fail due to this incorrectness. 
	However, this possibility was not supported during the interviews.
	
	\subsection{Systematic testing}
	Claim: \textit{I test a program by systematically checking all possible input values.}
	
	Answer: Average 3.67, standard deviation 0.75.
	This is similar to claim A.4 from Kolikant~\cite{Kolikant-2005}.
	In that study, 71\% of the high school students and 75\% of the college students agreed.
	In our population, 79\% agreed.
	This is a remarkable claim, because the exercise results showed that the students produced a very limited set of test cases that certainly did not cover all possibilities.
	
	\subsection{Overlooking cases}
	Claim: \textit{There is always the possibility that the program fails for some input value I have not discovered.}
	
	Answer: Average 4.59, standard deviation 0.62. This is similar to claim A.5 from Kolikant~\cite{Kolikant-2005}. In that case, 54\% of the high school students and 81\% of the college students agreed. In our population no less than 93\% agreed. This shows that the optimism exhibited in the previous section should not be taken too literally. Kolikant~\cite{Kolikant-2005} concludes from these numbers that students tend to describe their non-systematic methods as systematic. Our results strongly confirm this conclusion.
	
	\subsection{Time use}
	Claim: \textit{The ratio of time spent on programming and testing should be (1) 100/0, (2) 75/25, (3) 50/50, (4) 25/75, (5) 0/100.}
	
	Answer: Average 2.64, standard deviation 0.75. Of the respondents 48\% thought most of the time should be spent on programming, 41\% thought you should spend the same amount of time for both, and only 10\% thought that you should spend more time on testing. One of these students had the answer 0\% programming, 100\% testing, which is a strange answer.
	
	\subsection{Boundary values}
	Claim: \textit{In selecting test cases I take boundary values into account.}
	
	Answer: Average 3.53, standard deviation 0.85.
	However, it was established in the interviews that not all students know the meaning of the term boundary values.
	In the exercise results, we see boundary values used only occasionally.
	
	\section{Results - Interviews}\label{subsec:analysis_of_interviews}
	
	We interviewed eleven students.
	Here, we present our findings and illustrate them using  phrases from the transcribed interview texts.
	
	\subsection{Test cases are based on code inspection}
	Code inspection is very often mentioned as an approach to compose test cases.
	The functionality of the code is determined based on the code itself instead of on specification.
	
	\begin{iquote}`First, I've read the description of the exercise, after which I read the code thoroughly to determine its functionality. Otherwise, I am not able to determine the expected outcomes.'\end{iquote}
	
	\noindent
	Other examples of students indicating explicitly that they need the code to understand its functionality are:  
	
	\begin{iquote}`I read the text above the code and looked at the code to determine if I understood what happens in the code.'\end{iquote}   
	
	\begin{iquote}`Here, I read the code and hope to get more information on how it should work.'\end{iquote}  

	\noindent
	That students need the code to compose test cases is shown by the following examples:
	
	\begin{iquote}`I was surprised that there was no code! That means that you have to think about test cases based only on the specification!'\end{iquote}
	
	\begin{iquote}`There was no code available, so I have to think about how it works. So I imagined how it could be implemented, to see how it should work.'\end{iquote}
	
	\subsection{When a bug is found, a test case for that bug is composed}
	
	Many students mention that they are looking for bugs in the code. 
	For each bug that they find, they compose a test case.
	An example is:
	
	\begin{iquote}`Interviewer: And suddenly, you saw the error in the code?\\
		Student:
		Yes, and then I thought, I write \textsc{[1,2,3]} and then it is ready, on to the next one.'\end{iquote}
	
	\noindent
	Furthermore, students mention that, besides happy path testing,  the test cases are limited to the bugs found in the code. 
	This can explain the low number of test cases we observe in the students' solutions on the exercises (see Section \ref{sec:analysis_of_exercises}). 
	An example: 
	
	\begin{iquote}`Actually, I devise test cases more or less on what I see in the code, as if to say this is erroneous.'\end{iquote}
	
	Other methods of implementation based testing were never mentioned during the interviews.
	
	\subsection{Wrong test strategies}
	Wrong test strategies that are often mentioned are:
	random based test cases,
	happy path testing,
	pursue exhaustive testing,
	restrict test cases to the examples described as part of the exercise, 
	and restrict test cases to bugs found in the code.
	
	An example of random testing is the following. 
	On the question of whether the number of test cases that is sufficient, the student answers:
	\begin{iquote}
		`Student:
		For this, it is enough. \\
		Interviewer:
		You took some numbers, randomly, and  looked ...\\
		Student:
		... if they are correct. Yes.'\end{iquote}
	
	An example of happy path testing is:
	\begin{iquote}`It was more about figuring out .... how much ..., how often the longest period of frost took place, say .... when the longest period of frost took place. For testing, you need only negative numbers. If there are no negative numbers, there is no longest period of frost. So ...'\end{iquote}
	
	Some students pursue exhaustive testing.
	\begin{iquote}`Only integers are allowed. Thus, in that case all possible integers as input until the computer is not able to process them. That should be a physical problem. Yes.'\end{iquote}
	
	Sometimes, students limit the test cases to the example(s) given in the exercise test. This often leads to happy path testing too.
	\begin{iquote}`I used exactly the same examples as given in the exercise text.%
		'\end{iquote}
	\noindent	
	Finally, as we have mentioned earlier, students often limit test cases to bugs find in the code. 
	
	\subsection{Unnecessary or even impossible testing}
	
	Some students mention they add test cases to check value types although the program was coded in Java, which means the compiler detects type errors directly. We consider this as a misconception. 

	\begin{iquote}`Here I can add some characters and look how the program reacts because the program expects integers, but if I put in characters, then the program should chuck them out.'\end{iquote} 
	
	\begin{iquote}`If I have to input a number, then I input a string as for example \textsc{'hello'} and see what happens.'\end{iquote} 
	\noindent
	 
	Another misconception is that some students consider testing as a way of finding syntax errors.
	\begin{iquote}`It is possible that you forget a semicolon, and yet it does not work.
		In such a case it is good to look at each line of code and to see where it goes wrong.
		This is a way of testing.'\end{iquote}
	\noindent
	These misconceptions probably show students do not understand what a compiler does.
	
	\subsection{Lack of motivation}
	Students often do not see the necessity to test code thoroughly:
	\begin{iquote}`Student:
		No, if I had a computer, then I should apply much longer test cases. \\
		Interviewer: 
		Is there a reason you did not do that? \\
		Student:
		Yes, too much effort.'\end{iquote}
	
	The following example shows the importance of grading:
	\begin{iquote}`I think that, how important is the exercise ..., if it is for grading, then I should perform testing more elaborately then just looking at the code. That is possible,  then it works, but in cases of grading, then you should find all errors in the code.'\end{iquote}	
	
	The following example is related to attitude/engagement:
	\begin{iquote}`I do not find myself good. It was early in the morning. Is possible that I missed some things. The attitude I made the exercise with played a role too. For me, this research is not important, it is not my research.'\end{iquote}
	
	\subsection{Reading someone else's code is difficult}
	Students often mentioned that reading someone else's code is difficult. 
	\begin{iquote}`I experienced a lot of problems with the code conventions because I am used to place the brackets in a different way.'\end{iquote}
	\begin{iquote}`I did not understand the code really, because of course it is not my own code.'\end{iquote}
	\noindent
	This could be a reason for the few test cases students wrote.
	
	\subsection{Pen and paper versus working on a computer}
	Some students explicitly mentioned  that they prefer working on a computer instead of working with pen and paper. Working on a computer means running the code to see if it `works'.  
	\begin{iquote}`It is difficult for me to do it just with pen and paper. It is easier to do it on a computer. Then, you can easily see what happens while running the code, what the code does exactly.'\end{iquote}
	
	Students look for bugs by experimenting with the code, for which a computer is needed. 
	With pen and paper, this approach is not possible. 
	In fact, they debug the code instead of test it.
	This is an educational issue: students have to learn the differences between debugging and testing and have to learn how to write specification-based tests, probably the best with pen and paper. 
	
	\section{Conclusions}~\label{sec:conclusions_and_future_work} 
	Our long term goal is to improve the quality of the code that students produce,  through better testing education.
	To improve our test education, we need insight into student's misconceptions and their view on testing before they have had any relevant instruction about this topic.
	
	\subsection{Findings}
	
	\subsubsection{Test cases are based on code inspection}
	It was remarkable that students based their tests on code inspection, even in the case of an exercise with only a specification.
	For this exercise, they first thought about the code they would write to solve the problem.
	Some students could not write tests at all for some exercises because they did not understand the code.
	
	Conclusion: students at this level do not have the notion of basing tests on the specification.
	
	\subsubsection{Test cases for a bug}
	During the interviews and in the test cases defined during the exercises, we see that students read the code, find a bug and write a test case for that bug.
	This can partially be attributed to the test setup.
	During the exercises, students were asked if they think the code was correct and to write a test case that shows the bug if they believe the code was incorrect.
	This question can be a trigger to specifically search for bugs.
	
	However, it is apparent that many students moved on to the next exercise after designing a test for the (presumed) bug in the code.
	They did not take the time to think of other test cases.
	
	This strong focus on the given code shows that many of the students do not write test cases as a way to assure the correctness of a program during its complete life cycle, but more of a way to debug the code.
	This is consistent with the findings of Edwards regarding a trial-and-error approach to software development and testing~\cite{Edwards-2014}. 
	It is known from the literature that beginning students do not see a difference between testing and debugging~\cite{murphy2008debugging}.
	
	\subsubsection{Lack of systematic testing}
	Both the interviews and the exercises show that the students tend to limit themselves to `happy path testing'.
	This finding fits with the survey results showing that students are optimistic about the correctness of their code.	
	This is a known phenomenon~\cite{Edwards-2014}. 
	
	In the classification of Michaeli~\cite{michaeli2017addressing}, our students have a `level 1' understanding of software quality (thinking that software that successfully processes sample data works). 
	In the classification of Beizer~\cite{Beizer-Techniques} they are in phase 1 (thinking that the purpose of testing is to show that the software works) and in some cases phase 2 (thinking that the purpose of testing is to show that the software does not work).
	
	A more extreme misconception was found where students did not think at all about providing test cases, but merely copied the examples that were mentioned in the exercise text for the purpose of illustrating and clarifying the specification. This may be ascribed to misunderstanding the task.
	
	\subsubsection{Incomplete test sets}
	The exercises reveal that almost all the test sets defined by the students are far from complete, mostly only containing happy path test cases. 
	Specification-based requirements (such as robustness), as well as implementation-based requirements (such as coverage ratios) are not satisfied.
	The above findings explain these incomplete test sets well.      
	
	\subsubsection{Wrong test strategies}
	Besides happy path testing, we observed test cases restricted only to the examples as part of the exercise, and test cases restricted to bugs found in code.  
	Another remarkable test strategy we observed is exhaustive testing, i.e. trying to feed a function with all possible inputs.
	This is a known misconception: Complete Testing is Possible\footnote{See \url{https://www.tutorialspoint.com/software_testing/software_testing_myths.htm}}.
	These students described this approach, but did not try to show their test cases.
	One student mentioned the possibility of physical problems.
	
	\subsubsection{Lack of motivation}
	Many students showed a lack of motivation for testing.
	They are optimistic about the correctness of their own code and consider testing merely an additional burden.
	One reason may be that the test tasks are experienced as too simple to justify the extra work~\cite{Scatalon-2017}, while code inspection is still feasible.
	This leads to a paradox in testing education. 
	If the code is small enough to understand, testing is not a necessity.
	If the code becomes larger, students are unable to comprehend the code and are therefore unable to design tests (at least, white-box tests).
	
	\subsubsection{Time spent to test}
	The time spent by students to read an exercise, to define test cases and to inspect the code is remarkably short.
	This observation matches the findings of happy path testing, test cases based on code inspection specifying a test case for found bugs, as well as a lack of motivation. 

	\subsubsection{Unnecessary or even impossible testing}
	Although the language we use is Java, some students proposed type testing in their answers.
	Possibly, students tested the program on robustness, i.e. how it reacts to erroneous inputs.
	Also, some students used testing as a way to find syntax errors. 
	Because type checking and syntax checking are performed by the compiler, we consider these as misconceptions, i.e. unnecessary testing.
	We did not find this type of misconception in existing research. 
	
	\subsection{Regarding Kolikant's findings}
	Regarding Kolikant's study~\cite{Kolikant-2005}, our population of students reveals more mistrust concerning the correctness of a program based on reasonable output of that program: 24\% of our population versus 50\% of the population of Kolikant consider reasonable output to be a sufficient indicator of correctness.
	The difference increases in case of complicated calculations: our population 3\% versus Kolikant 33\% in case of high school students and 69\% in case of college students.
	
	A similar observation was done involving the no-testing approach in the case that a programmer is certain that his/her program is correct.
	In our study, only 7\% agreed that testing is not necessary if the code compiles, where in the Kolikant study 31\% agreed with that statement.
	These findings follow from the pre-exercises surveys. 
	
	Almost similar to Kolikant, we observe that 79\% of the students think that they test systematically.
	The exercises and interviews show that they produced a very limited set of test cases that certainly did not cover all possibilities.
	We also, like Kolikant,  conclude that students tend to describe their non-systematic methods as systematic.
\bibliographystyle{IEEEtran}
\bibliography{test}
	
\end{document}